\documentclass[largeformat]{interact}
\usepackage{epstopdf}% To incorporate .eps illustrations using PDFLaTeX, etc.
\usepackage{subfigure}% Support for small, `sub' figures and tables

\usepackage{color}
\usepackage{graphicx}
\usepackage{epsf}
\usepackage{pdfpages}

\usepackage[numbers,sort&compress,merge]{natbib}% Citation support using natbib.sty
\bibpunct[, ]{(}{)}{,}{n}{,}{,}% Citation support using natbib.sty
\theoremstyle{plain}% Theorem-like structures

\theoremstyle{definition}

\theoremstyle{remark}

\begin{document}

%\articletype{ARTICLE TEMPLATE}

\title{A unified approach to $\Lambda$-, $\Xi$- and $V$-type systems with one continuum}

\author{
\name{Surajit Sen\textsuperscript{a}\thanks{CONTACT S.S. Author. Email: ssen55@yahoo.com} and Tushar Kanti Dey \textsuperscript{a} and Bimalendu Deb\textsuperscript{c}}
\affil{\textsuperscript{a,}\textsuperscript{b}Department of Physics, Guru Charan College, Silchar 788004, India;\\ \textsuperscript{c}Department of Materials Science, Raman Center for Atomic, Molecular and Optical Sciences, Indian Association for the Cultivation of Science, Jadavpur, Kolkata 700032, India}
}

\maketitle

\begin{abstract}
We present a systematic approach to classify the three-level-like models with  two  bound states coupled to a continuum. It is shown that, when one of the discrete levels of usual three-level Lambda- ($\Lambda$), cascade ($\Xi$) or Vee ($V$)-type systems is replaced by a continuum of states, the resulting each model can be classified into three distinct categories with nine possible configurations. We show that all these models are {\it exactly} solvable. We obtain and compare the asymmetric Fano line shapes of the spectra for all the models. Our results are important for exploring new coherent effects in a variety of physical systems involving continuum-bound coupling such as photoassociation of cold atoms, plasmonics, quantum dots, photonic crystals, electromagnetic metamaterials and so on.
\end{abstract}


\section{Introduction}
\par
The level-structure in atomic, molecular and nuclear physics is possibly the most important signature of quantum mechanics which time and again deserves exhaustive exploration. In these branches of physics, bound states always coexist with a continuum of states. Here the continuum refers to the states of the scattering between two particles. For instance, an ionisation continuum of an atom implies the scattering states between an ion and an electron or a dissociation continuum of a diatomic molecule means the collisional states of two atoms that constitute the molecule. So there are three classes of quantum-mechanical transitions known as, bound-bound, bound-continuum and continuum-continuum transitions.  Each class of these transitions has its own  characteristic line-shape. There are physical situations where different classes of transitions can happen simultaneously leading to quantum interference between the transition pathways. One such situation arises in the phenomenon of photoionization or autoionization of an atom showing an asymmetric line-shape. This feature is in sharp contrast to the typical symmetric Lorentzian pattern found in the bound-bound transition. The origin of this asymmetry, which is often observed in the optical spectroscopy, was understood after the pioneering work of Fano who first  pointed out the crucial role of the coupling between the continuum and a discrete bound state \cite{Fano1961}. He argued that the mixing of the bound state with apparently structureless continuum leads to the formation of an effective structured continuum known as `{\it autoionization state}'. The asymmetric Fano profile is an ubiquitous feature of many natural and artificially prepared materials like plasmonic nanoparticles \cite{Lukyanchuk2010}, quantum dots, photonic crystals and electromagnetic metamaterials, etc \cite{Khanikaev2013,Trocha2007, Ridolfo2010}. The theoretical framework of Fano ignores radiative dissipation or damping which results from the coupling of an excited atomic or molecular state with the surrounding environment of background electromagnetic modes of the quantum vacuum or zero-photon states of radiation, often referred to as the bath or reservoir system. Such system constitutes another kind of continuum which is quite distinct from the continuum of two-particle scattering states that we consider in this paper. Usually, radiative dissipation is treated by a weak-coupling system-reservoir theory within Born and Markoff approximations. In case of autoionisation, the neglect of radiative dissipation is justified because the electron is coupled to the ionisation continuum far too strongly compared to its coupling to the reservoir of electromagnetic modes.
\par
%\vspace{0.3cm}
%\rule[0.1cm]{16cm}{0.02cm} \\
A bound-continuum coupled system embedded in environment or in reservoir can be treated as an open quantum system using non-hermitian Hamiltonian. Many interesting effects of open quantum system, such as Fano-Feshbach resonance around the exceptional points has some special place in atomic and molecular physics \cite{Heiss2014,Eleuch2015}, dynamical phase transition (DPT), width bifurcation etc. \cite{Muradyan2002,Boutabba2009,Eleuch2013}. Both Fano and non-hermitian Hamiltonian methods are capable of taking into account the strong-couplings of the multilevel systems with the continuum, while the master equation approach based on Born and Markoff approximations is applicable only in the weak-coupling limit. In Feshbach projection operator method \cite{Feshbach1958}, the effective Hamiltonian of the sub-system essentially turns  into a non-hermitian operator. Using such an effective Hamiltonian approach, the formation of a bound state in the continuum \cite{Neumann1929} in the cold collisions of atoms has been predicted \cite{Deb2014} recently. While the open quantum system approach is applicable to a variety of physical problems including large and many-body quantum systems, such as Dicke superradiance \cite{Dicke1954}, Fano method seems to be more suitable for describing a few level system coupled to one (or two) continuum (continua). Fano method has unique feature: It can treat both the bound and continuum states on an equal footing. So, in the continuum-bound coupled systems where bound-continuum coherence is important, continuum can be an integral part of coherent phenomena or dynamics in this method. For instance, atom-molecule coherence in photo- or magneto-association of cold atoms \cite{Kohler2006} and Rabi oscillations between atomic and molecular states \cite{Saha2016} arise due to continuum-bound coherence.

%\vspace{0.3cm}
%\rule[0.1cm]{16cm}{0.02cm} \\

\par
Fano method can describe exactly the quantum states of systems where not only a single bound state interacts with a continuum, but also of multiple interacting or non-interacting bound states become coupled to a single continuum or multiple continua \cite{Deng1984,Paspalakis1999, Glutsch2002}. Apart from autoionization or photoionization systems where a structured continuum of quantum states appears quite naturally, it is possible to induce a structure in an otherwise structureless continuum by lasers leading to laser-induced continuum structures (LICS) \cite{Shao1991, Cavalieri1995, Zhang1992, Shapiro2006, Rangelov2007, Dalton1994}. This opens up immense possibility of coherent manipulation of continuum-bound coupled systems. A minimal extension of Fano model is  three-level-like models involving one continuum and two bound states forming $\Lambda$- $V$- and $\Xi$-type systems as depicted in Figs.1-3.

\begin{figure}[ht]
\centering
\subfigure[]
{
\includegraphics[width=4.00cm,height=4.25cm,keepaspectratio]{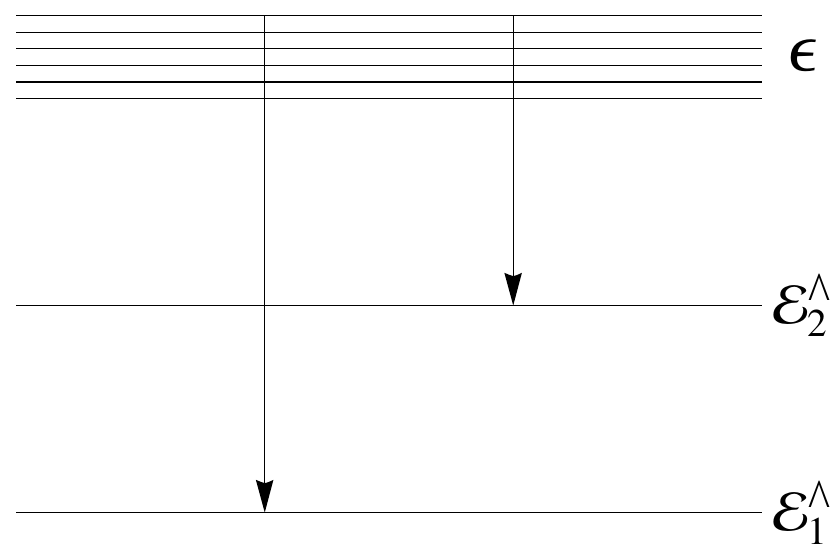}
\label{subfigure1a}}
\quad
\subfigure[]{
\includegraphics[width=4.00cm,height=4.25cm,keepaspectratio]{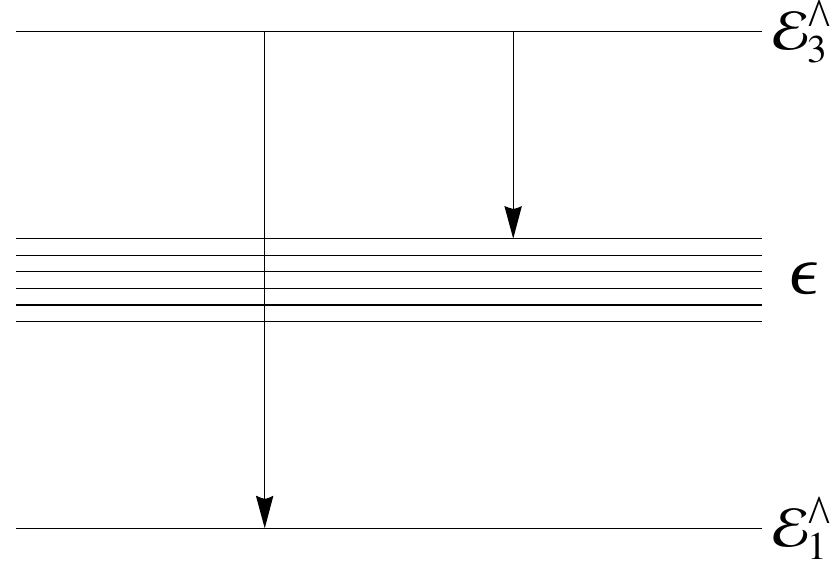}
\label{subfigure1b}}
\subfigure[]{
\includegraphics[width=4.00cm,height=4.25cm,keepaspectratio]{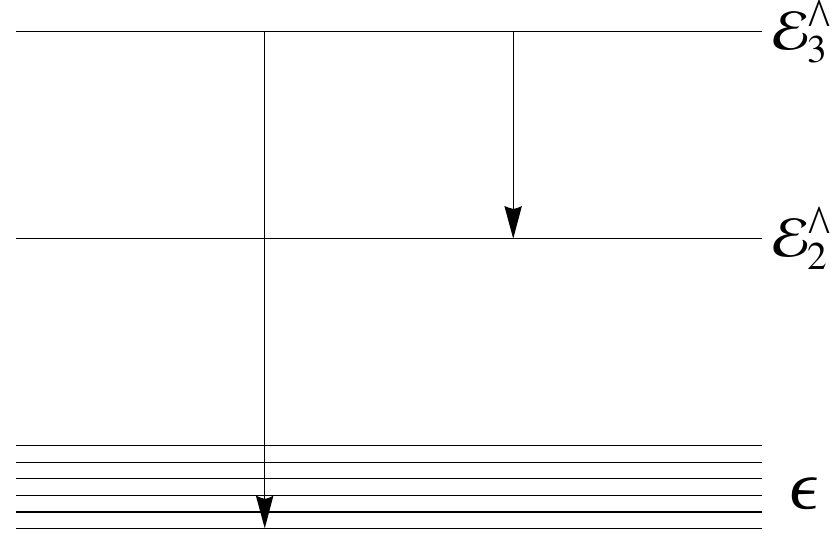}
\label{subfigure1c}}
\caption{ Lambda-like system with  upper (a), middle (b) and lower (c) level as continuum state.}
\label{figure1}
\end{figure}
In quantum optics, the three-level systems with three discrete bound states exhibit a rich class of coherent phenomena such as  two-photon coherence, double-resonance,  resonance Raman scattering, coherent population trapping, STIRAP, quantum jumps, quantum Zeno effects, electromagnetically induced transparency (EIT) etc. \cite{Sen2012,Yoo1985}.
\par
In contrast to discrete thee-level models, three-level-like continuum models have not attracted much research interests in the context of coherent phenomenon. Because,  generally accepted notion  is that whenever a continuum of states is coupled to a bound state, there is irreversible decay of the probability of the bound-state into the continuum.  However,  Fano model is known to give rise to a confluence of coherences \cite{Rzazewski1981} and therefore its extension to three-level-like systems is expected to give rise to coherent phenomena. The underlying assumption for the validity of the Fano model  is that the bound states and the continua are non-decaying. In practice, this implies that radiative as well as  non-radiative decay of the bound-continuum coupled systems should be negligible. In other words, the continuum-bound coupling should be much stronger than the damping constants of the  bound or continuum states. In this work, we consider strong-coupling models where the radiative environment effects can be negligible. Under such conditions, three-level-like systems with a continuum are expected to provide a new avenue for hither-to-unexplored coherent effects. It is therefore of interest to develop a unified approach towards the  Fano-inspired  three-level-like models with one continuum.

\begin{figure}[ht]
\centering
\subfigure[]{
\includegraphics[width=4.00cm,height=4.25cm,keepaspectratio]{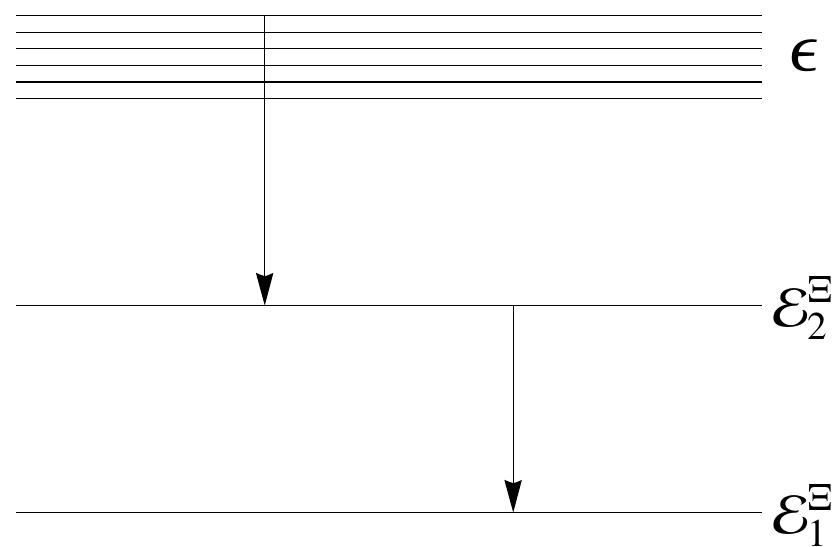}
\label{subfigure2a}}
\quad
\subfigure[]{
\includegraphics[width=4.00cm,height=4.25cm,keepaspectratio]{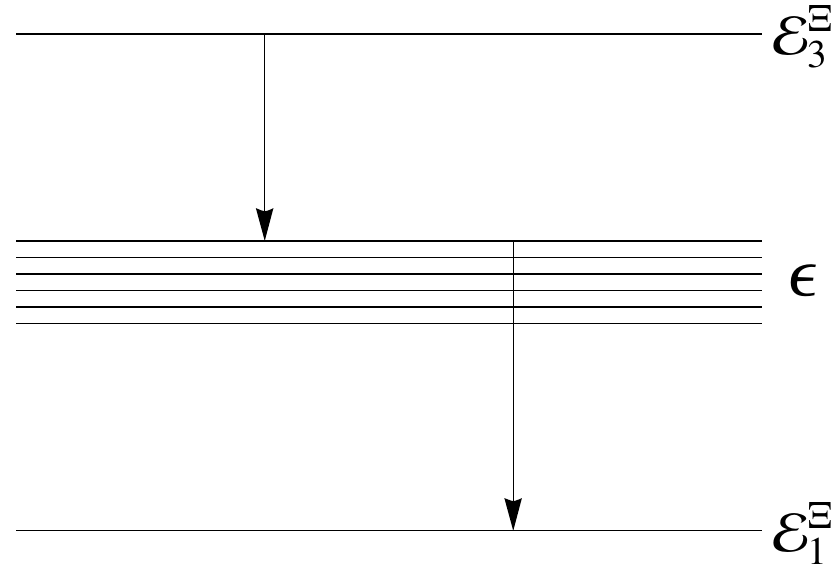}
\label{subfigure2b}}
\subfigure[]{
\includegraphics[width=4.00cm,height=4.25cm,keepaspectratio]{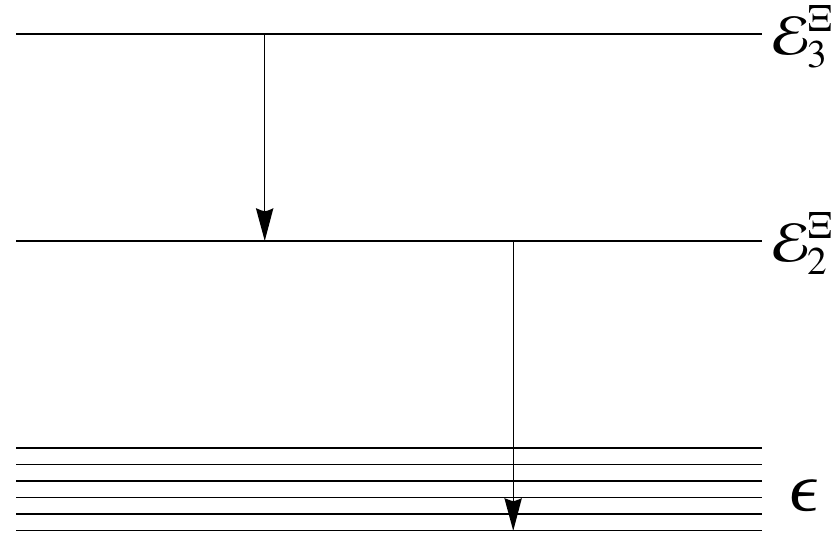}
\label{subfigure2c}}
\caption{Cascade-like system with upper (a), middle (b) and lower (c) level as continuum state.}
\label{figure2}
\end{figure}
\par
In the recent past, dissipation has been shown to be treatable within the framework of the Fano model \cite{Agarwal1984a,Agarwal1984b, Agarwal2013}. This opens up a new prospect of controlling dissipation by coherent manipulation of level structures in the systems with a continuum, providing a new perspective in the dark resonance \cite{Cohen-Tannoudji1990} and related phenomena like EIT and STIRAP. The EIT effect is connected with a wide verity of quantum-optical phenomena such as, manipulation of the dispersive property of a medium \cite{Suominen1991, Boller1991,Kasapi1995,Harris1997}, slow light \cite{Hau1999,Dutton2004}, quantum switching \cite{Li2012}, and many other effects \cite{Marangos1998,Fleischhauer2005, Sen2015}. Recently, some endeavors have been made to realize similar effects in three-level like systems with a continuum of states \cite{Carroll1992,Carroll1993,Nakajima1994,Unanyan1998,Vardi1999,Vardi1997}. Particularly, with the recent advent of continuum-bound photoassociation (PA) spectroscopy \cite{Weiner1999,Jones2006} by which two colliding cold atoms are associated into a molecule via a single-photon absorption, coherent effects in three-level-like PA systems have become important \cite{Vardi1997,Saha2016,Sardar2016}. In fact, EIT-like spectral features in two-color PA has been experimentally observed\cite{Schloder2003, Winkler2005, Jones1997,Sardar2016}. Furthermore, the bound-continuum dipole coupling due to a  control field is shown to induce an asymmetric pattern in the dispersion profile of the probe field \cite{Quoc2012, Dinh2013, Raczynski2006, Agarwal2013, Qu2013}. The $\Lambda$-type three-level atomic medium in the EIT regime is important for soliton propagation \cite{Eleuch2004,Boutabba2009} and ultrafast Raman adiabatic passage \cite{Dridi2009}. The transition from Autler-Townes splitting (ATS) \cite{Autler1955} to EIT \cite{Anisimov2011}has been shown to exhibit features such as avoided crossing and width bifurcation \cite{Giner2013,Abdumalikov2010} that usually pertain to an open quantum system \cite{Rotter2015}. This indicates that a three-level like continuum system will provide new insight into ATS to EIT transitions.

\begin{figure}[ht]
\centering
\subfigure[]{
\includegraphics[width=4.00cm,height=4.25cm,keepaspectratio]{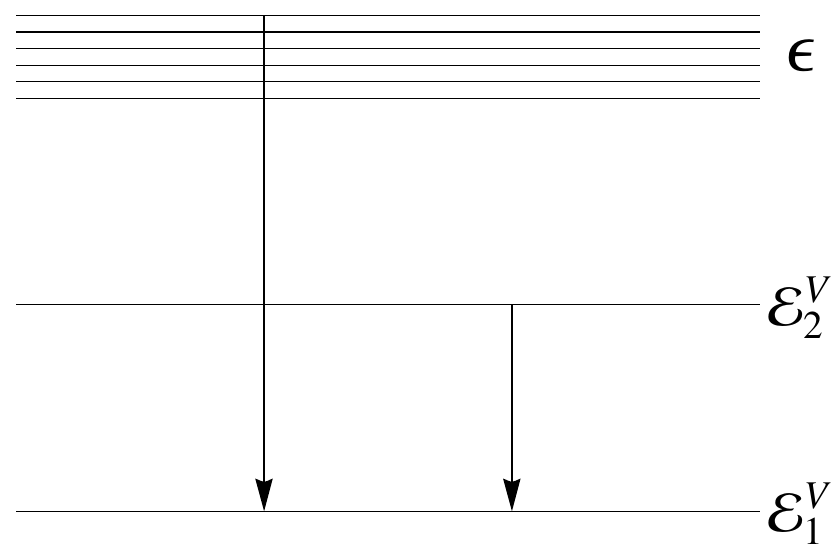}
\label{subfigure3a}}
\quad
\subfigure[]{
\includegraphics[width=4.00cm,height=4.25cm,keepaspectratio]{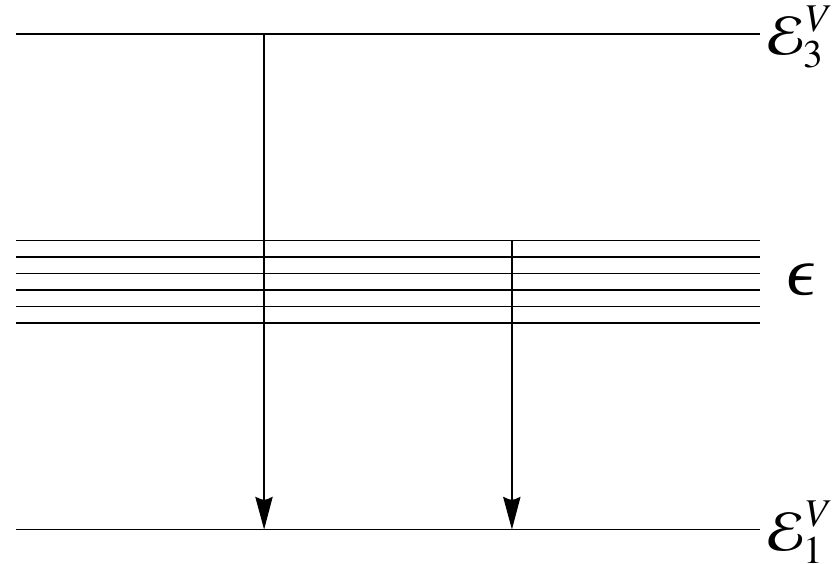}
\label{subfigure3b}}
\subfigure[]{
\includegraphics[width=4.00cm,height=4.25cm,keepaspectratio]{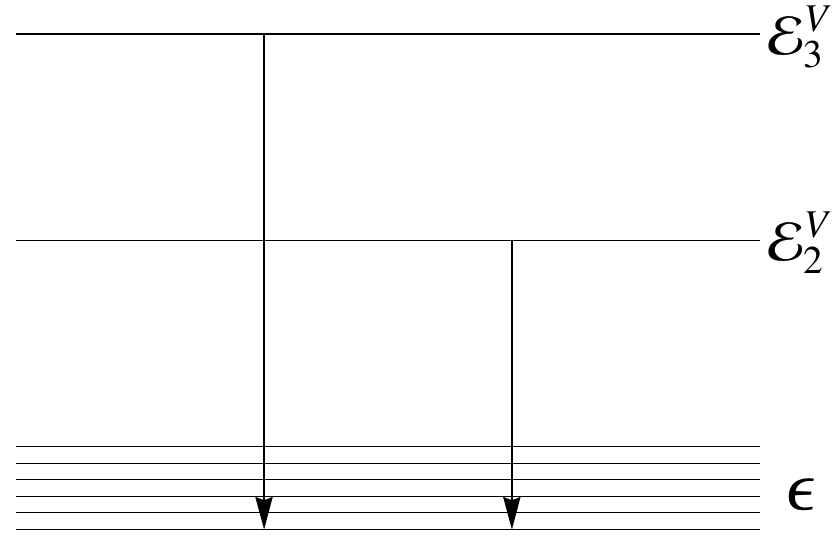}
\label{subfigure3c}}
\caption{Vee-like system with upper (a), middle (b) and lower (c) level as continuum state.}
\label{figure3}
\end{figure}

\par
The remaining sections of the paper are organized as follows: In Section II we discuss possible classification of the three-level like systems with one continuum and two distinct bound states and develop Hamiltonians of such configurations. In Section III, we present the detailed derivation of the $\Lambda$-like system and discuss the characteristics of its line-shapes. The derivation of the line-shapes of the remaining $\Xi$ and $V$ systems are similar as in the $\Lambda$ system. The results of $\Xi$ and $V$ systems are presented in Section IV and V. In Section VI, we compare the line-shapes of each system and point out their crucial differences. In the concluding section, we summarize our results and discuss the outlook of our work.
\section{The models}
\par
For three different positions of the continuum state, namely, the upper (U), middle (M)and lower (L) levels, we have three possible configurations, depending on the two transitions, as shown in Figs.1-3. We note that, for each of the $\Lambda$, $\Xi$ and $V$-like systems, we have one configuration with two bound-continuum transitions as shown in Figs.1(a), 2(b) and 3(c), while for the remaining pair of configurations, namely, Figs.1(b, c), 2(a, c) and 3(a, b), we have one bound-continuum and one bound-bound transition, respectively. The Hamiltonians of these three types of $\Lambda$ system are given by
\begin{subequations}\label{one}
\begin{align}\label{onea}
H^{\Lambda}_U = \sum_{n=1}^{2}{\varepsilon_{n}^\Lambda}|n\rangle \langle n| + \int d{\epsilon} \epsilon |3,\epsilon\rangle \langle\epsilon,3|+
+\sum_{n=1}^{2}\int d{\epsilon}V_{n3}^\Lambda(\epsilon)|3,\epsilon\rangle \langle n|+h.c.,
\end{align}
for Fig.~\ref{subfigure1a},
\begin{align}\label{oneb}
H^{\Lambda}_M = {\varepsilon_{1}^\Lambda}|1\rangle \langle1|+ \int d{\epsilon} \epsilon |2,\epsilon\rangle \langle\epsilon,2|+{\varepsilon_{3}^\Lambda}|3\rangle \langle3|
+g_{13}^\Lambda|1\rangle \langle3|+\int d{\epsilon}V_{32}^\Lambda(\epsilon)|2,\epsilon\rangle \langle3|+h.c.,
\end{align}
for Fig.~\ref{subfigure1b},
\begin{align}\label{onec}
H^{\Lambda}_L = \int d{\epsilon} \epsilon |1,\epsilon\rangle \langle\epsilon,1|+{\varepsilon_{2}^\Lambda}|2\rangle \langle2|+ {\varepsilon_{3}^\Lambda}|3\rangle \langle3|
+g_{23}^\Lambda|2\rangle \langle3|+\int d{\epsilon}V_{31}^\Lambda(\epsilon)|1,\epsilon\rangle \langle3|+h.c.,
\end{align}
\end{subequations}
for Fig.~\ref{subfigure1c}. Here, $\varepsilon^\Lambda_i$, $g_{ij}$  and $V_{ij}^\Lambda$ ($i=1,2,3$ with $i\neq j$)) are the energy of the bound state, bound-bound and the bound-continuum coupling constants, respectively, with $|n\rangle (n = 1, 2, 3 )$ being a bound state.
\noindent
Similarly, the Hamiltonians of the $\Xi$ model are given by,
\begin{subequations}\label{two}
\begin{align}\label{twoa}
H^{\Xi}_U = \sum_{n=1}^{2}{\varepsilon_{n}^\Xi}|n\rangle \langle n| + \int d{\epsilon} \epsilon |3,\epsilon\rangle \langle\epsilon,3|
+g_{12}^\Xi|1\rangle \langle2|+\int d{\epsilon}V_{23}^\Xi(\epsilon)|2\rangle \langle\epsilon,3|+h.c.,
\end{align}
for Fig.~\ref{subfigure2a},
\begin{align}\label{twob}
H^{\Xi}_M = {\varepsilon_{1}^\Lambda}|1\rangle \langle1|+ \int d{\epsilon} \epsilon |2,\epsilon\rangle \langle\epsilon,2|+{\varepsilon_{3}^\Xi}|3\rangle \langle3|
+\int d{\epsilon}V_{12}^\Xi(\epsilon)|2,\epsilon\rangle \langle3|+\int d{\epsilon}V_{32}^\Xi(\epsilon)|2,\epsilon\rangle \langle3|+h.c.,
\end{align}
for Fig.~\ref{subfigure2b},
\begin{align}\label{twoc}
H^{\Xi}_L = \int d{\epsilon} \epsilon |1,\epsilon\rangle \langle\epsilon,1|+{\varepsilon_{2}^\Xi}|2\rangle \langle2|+{\varepsilon_{3}^\Xi}|3\rangle \langle3|
+\int d{\epsilon}V_{21}^\Xi(\epsilon)|1,\epsilon\rangle \langle2|+g_{23}^\Xi|2\rangle \langle3|+h.c.,
\end{align}
\end{subequations}
for Fig.~\ref{subfigure2c}. Proceeding in the same way, the Hamiltonians of the $V$ model are given by,
\begin{subequations}\label{three}
\begin{align} \label{threea}
H^{V}_U = \sum_{n=1}^{2}{\varepsilon_{n}^V}|n\rangle \langle n| + \int d{\epsilon} \epsilon |3,\epsilon\rangle \langle\epsilon,3|
+g_{12}^\Xi|1\rangle \langle2|+\int d{\epsilon}V_{13}^V(\epsilon)|1\rangle \langle\epsilon,3|+h.c.,
\end{align}
for Fig.~\ref{subfigure3a},
\begin{align}\label{threeb}
H^{V}_M = {\varepsilon_{1}^\Lambda}|1\rangle \langle1|+ \int d{\epsilon} \epsilon |2,\epsilon\rangle \langle\epsilon,2|+{\varepsilon_{3}^V}|3\rangle \langle3|
+g_{13}^\Xi|1\rangle \langle3|+\int d{\epsilon}V_{12}^V(\epsilon)|2,\epsilon\rangle \langle1|+h.c.,
\end{align}
for Fig.~\ref{subfigure3b},
\begin{align}\label{threec}
H^{V}_L = \int d{\epsilon} \epsilon |1,\epsilon\rangle \langle\epsilon,1|+{\varepsilon_{2}^V}|2\rangle \langle2|+{\varepsilon_{3}^V}|3\rangle \langle3|
+\int d{\epsilon}V_{21}^V(\epsilon)|1,\epsilon\rangle \langle2|+\int d{\epsilon}V_{31}^\Xi(\epsilon)|1,\epsilon\rangle \langle3|+h.c.,
\end{align}
\end{subequations}
for Fig.~\ref{subfigure3c}.
\par
The time-independent Schr\"{o}dinger equation  of a generic  system ($i=\Lambda, \Xi, V$) is given by
\begin{equation}\label{four}
H^{i}|\psi^i(\varepsilon)\rangle=\varepsilon^i|\psi^i(\varepsilon)\rangle,
\end{equation}
where $\varepsilon^i$ is the energy eigenvalue with the wave function given by
\begin{subequations}\label{five}
\begin{align}\label{fivea}
|\psi^i(\varepsilon)\rangle = a^i(\varepsilon)|1\rangle +b^i(\varepsilon)|2\rangle + \int d{\epsilon}c^i(\epsilon,\varepsilon)|3,\epsilon \rangle,
\end{align}
for Fig.~\ref{subfigure1a}, Fig.~\ref{subfigure2a} and Fig.~\ref{subfigure3a},
\begin{align}\label{fiveb}
|\psi^i(\varepsilon)\rangle = a^i(\varepsilon)|1 \rangle+\int d{\epsilon}b^i(\epsilon,\varepsilon)|2,\epsilon \rangle +c^i(\varepsilon)|3 \rangle,
\end{align}
for Fig.~\ref{subfigure1b}, Fig.~\ref{subfigure2b} and Fig.~\ref{subfigure3b},
\begin{align}\label{fivec}
|\psi^i(\varepsilon)\rangle = \int d{\epsilon}a^i(\epsilon,\varepsilon)|1,\epsilon\rangle+b^i(\varepsilon)|2\rangle + c^i(\varepsilon)|3\rangle,
\end{align}
\end{subequations}
for Fig.~\ref{subfigure1c}, Fig.~\ref{subfigure2c} and Fig.~\ref{subfigure3c}, respectively. The amplitudes of three levels $a^{i}$, $b^{i}$ and $c^{i}$ appearing in Eq.\eqref{five} are energy-normalized as
\begin{subequations}\label{six}
\begin{equation} \label{sixa}
a^{i}(\varepsilon){a^{i}({\varepsilon}')}^*+b^{i}(\varepsilon){b^{i}({\varepsilon}')}^*
+\int c^{i}(\varepsilon,\epsilon) {c^{i}(\varepsilon',\epsilon)}^* d{\epsilon} = \delta(\varepsilon^i-{\varepsilon^i}'),
\end{equation}
\begin{equation}\label{sixb}
a^{i}(\varepsilon){a^{i}({\varepsilon}')}^*+ \int b^{i}(\varepsilon',\epsilon)(\varepsilon){b^{i}({\varepsilon',\epsilon})}^*
+ c^{i}(\varepsilon){c^{i}(\varepsilon)}^* d{\epsilon} = \delta(\varepsilon^i-{\varepsilon^i}').
\end{equation}
\begin{equation}\label{sixc}
\int a^{i}(\varepsilon,\epsilon) {a^{i}(\varepsilon',\epsilon)}^* d{\epsilon}+b^{i}(\varepsilon){b^{i}({\varepsilon}')}^*
+c^{i}(\varepsilon){c^{i}({\varepsilon}')}^* = \delta(\varepsilon^i-{\varepsilon^i}'),
\end{equation}
\end{subequations}
respectively. Having classifying the models of all configurations, we proceed to evaluate the amplitudes of the bound states for all
the systems and their respective line shapes.
\par
Note that, in writing these Hamiltonians we have not assumed the nature of interactions (or couplings) between bound and
continuum states or between two bound states. So, our models are quite general in the sense that these interactions can be
induced by applying external dynamic fields, for instance in the case of optical dipole transitions one can apply two laser fields in an appropriate configuration of a physical system with a continuum, or in the case of magnetic transitions one can realise such model by applying oscillating magnetic or radio-frequency fields. Since the interaction part of our Hamiltonians are taken to be time-independent, these Hamiltonians represent effective Hamiltonians under rotating-wave approximations where the harmonically varying time-dependence of the interaction part can be eliminated by a unitary transformation.

\vspace {.5cm}
\par
\section{\label{3} $\Lambda$ system}
\subsection{\label{3a} $\Lambda$ system with upper state being continuum}
\par
For the $\Lambda$ system ($i=\Lambda$) shown in Fig.~\ref{subfigure1a}, plugging  Eq.\eqref{one} and Eq.\eqref{fivea} in Eq.\eqref{four} and then taking the inner product,  we obtain, (We drop the symbol $\Lambda$ for notational convenience),
\begin{subequations}\label{seven}
\begin{equation}\label{sevena}
c(\epsilon,\varepsilon)(\epsilon-\varepsilon) + a(\varepsilon)V_{13}(\epsilon)+b(\varepsilon)V_{23}(\epsilon)=0,
\end{equation}
\begin{equation}\label{sevenb}
b(\varepsilon)(\varepsilon_2-\epsilon) + \int d \epsilon c(\epsilon,\varepsilon)V_{23}(\epsilon)=0,
%\Rightarrow b(\epsilon)=\frac{\int d \epsilon c(\epsilon)V_{\epsilon 23}(\epsilon)}{\epsilon_2-\epsilon^\Lambda}\\
\end{equation}
\begin{equation}\label{sevenc}
a(\varepsilon)(\varepsilon_1-\epsilon) + \int d \epsilon c(\epsilon,\varepsilon)V_{13}(\epsilon)=0,
%\Rightarrow c(\epsilon)=\frac{\int d \epsilon c(\epsilon)V_{\epsilon 13}(\epsilon)}{\epsilon_1-\epsilon^\Lambda}
\end{equation}
\end{subequations}
respectively. Substituting the value of $c(\epsilon,\varepsilon)$ from Eq.\eqref{sevena} into Eqs.\eqref{sevenb} and \eqref{sevenc} we have,
\begin{subequations}\label{eight}
\begin{equation}\label{eighta}
\int d\epsilon \frac{V_{23}(\epsilon)\left(a(\varepsilon) V_{13}(\epsilon)+b(\varepsilon) V_{23}(\epsilon)\right)}{\varepsilon-\epsilon}
+b(\varepsilon) \left(\epsilon_2-\varepsilon\right)=0
\end{equation}
\begin{equation}\label{eightb}
\int d\epsilon \frac{V_{13}(\epsilon)\left(a(\varepsilon) V_{13}(\epsilon)+b(\varepsilon)V_{23}(\epsilon)\right)}{\varepsilon-\epsilon } +b(\varepsilon)\left(\epsilon_1-\varepsilon\right)=0.
\end{equation}
\end{subequations}
To separate the principal part we use Dirac prescription, namely,
\begin{equation}\label{nine}
\frac{1}{\varepsilon-\epsilon }\to \frac{\mathbb{P}}{\varepsilon-\epsilon }+Z(\varepsilon)\delta(\varepsilon-\epsilon),
\end{equation}
and Eqs.\eqref{eight} can be written as,
\begin{subequations}\label{ten}
\begin{equation}\label{ten1}
b(\varepsilon) \left(\varepsilon-\varepsilon_2\right)+
Z(\varepsilon)V_{1(3\epsilon)}\big(a(\varepsilon) V_{1(3\epsilon)}+b(\varepsilon)V_{2(3\epsilon)}\big)=0,
\end{equation}
\begin{equation}\label{ten2}
a(\varepsilon) \left(\varepsilon-\varepsilon_1\right)+
Z(\varepsilon)V_{2(3\epsilon)}\big(a(\varepsilon) V_{1(3\epsilon)}+b(\varepsilon)V_{2(3\epsilon)}\big)=0.
\end{equation}
\end{subequations}
In deriving Eq.\eqref{ten}, we drop the principal part integral and assume that the interaction terms are independent of energy, i.e., $V_{13}(\epsilon)=V_{1(3\epsilon)}$ and $V_{23}(\epsilon)=V_{2(3\epsilon)}$. Eqs.\eqref{ten} readily gives
\begin{equation}\label{eleven}
Z(\varepsilon)=\frac{(\varepsilon - \epsilon_1)(\varepsilon - \epsilon_2)}{(\varepsilon - \epsilon_2)|V_{1(3\epsilon)}|^2+(\varepsilon - \epsilon_1)|V_{2(3\epsilon)}|^2}
\end{equation}
and the ratio of the amplitudes of two bound states are given by
\begin{equation}
\begin{aligned}\label{twelve}
%\begin{subequations}\label{twelve}
%\begin{equation}
\frac{a(\varepsilon)}{b(\varepsilon)}&=\frac{\varepsilon-\epsilon_2-Z(\varepsilon)|V_{2(3\epsilon)}|^2}{Z(\varepsilon)V_{1(3\epsilon)}V_{2(3\epsilon)}} \\
%\end{equation}
%\begin{equation}
&=\frac{(\varepsilon-\epsilon_2)V_{1(3\epsilon)}}{(\varepsilon-\epsilon_1)V_{2(3\epsilon)}}.
\end{aligned}
\end{equation}
%\end{subequations}
\par
To evaluate the amplitudes of the bound states, i.e., $a(\varepsilon)$ and $b(\varepsilon)$, we shall use the normalization condition given by Eq.\eqref{sixa}. Using Dirac prescription once again, Eq.\eqref{sevena} can be written as
\begin{equation}\label{thirteen}
c(\epsilon,\varepsilon) = (a(\varepsilon)V_{1(3\epsilon)}+b(\varepsilon)V_{2(3\epsilon)})\Big(\frac{1}{\varepsilon-\epsilon }+Z(\varepsilon)\delta(\varepsilon-\epsilon)\Big).
\end{equation}
Furthermore, using the identity,
\begin{equation} \label{fourteen}
%\begin{aligned}
\frac{1}{(\varepsilon'-\epsilon)(\varepsilon-\epsilon)} = \frac{\mathbb{P}}{(\varepsilon'-\varepsilon)}\Big(\frac{\mathbb{P}}{\varepsilon-\epsilon }-\frac{\mathbb{P}}{\varepsilon'-\epsilon }\Big)
+ \pi^2\delta(\varepsilon-\epsilon)\delta(\varepsilon'-\epsilon),
%\end{aligned}
\end{equation}
and once again neglecting the principal part, the probability density of the continuum state is given by
%\begin{widetext}
\begin{equation} \label{fifteen}
\begin{aligned}
&c(\epsilon,\varepsilon){c(\epsilon,\varepsilon')}^* =
\big(a(\varepsilon)V_{1(3\epsilon)}+b^{\Lambda}(\varepsilon)V_{2(3\epsilon)}\big)\big(a(\varepsilon')^*V_{1(3\epsilon)}+b(\varepsilon')^*V_{2(3\epsilon)}\big)\\
&\times\Big[(\pi^2+Z(\varepsilon)Z(\varepsilon') \delta(\varepsilon-\epsilon)\delta(\varepsilon'-\epsilon)+
\frac{1}{\varepsilon'-\epsilon }Z(\varepsilon)\delta(\varepsilon-\epsilon)+
\frac{1}{\varepsilon-\epsilon}Z(\varepsilon')\delta(\varepsilon'-\epsilon)\Big].
\end{aligned}
\end{equation}
Plugging back Eq.\eqref{fifteen} into Eq.\eqref{sixa} and integration over the continuum yields,
%\begin{subequations}\label{sixteen}
\begin{eqnarray} \label{sixteen}
& & a(\varepsilon){a(\varepsilon')}^{*}+b(\varepsilon){b(\varepsilon')}^{*}\nonumber \\
& & +
\frac{1}{\varepsilon'-\varepsilon}\big(a(\varepsilon)V_{1(3\epsilon)} +b(\varepsilon)V_{2(3\epsilon)}\big)  \big(a(\varepsilon')^{*}V_{1(3\epsilon)}+b(\varepsilon')^{*}V_{2(3\epsilon)}\big) \big(Z(\varepsilon)-Z(\varepsilon')\big) \\
& & + \big(a(\varepsilon)V_{1(3\epsilon)}+b(\varepsilon)V_{2(3\epsilon)}\big)\big(a(\varepsilon')^{*}V_{1(3\epsilon)}+
b(\varepsilon')^{*}V_{2(3\epsilon)}\big)\big(\pi^2+Z(\varepsilon)Z(\varepsilon')^{*}\big)
\delta(\varepsilon-\varepsilon')=\delta(\varepsilon-\varepsilon'). \nonumber
%\end{align}
%\end{equation}
%\begin{equation}
\end{eqnarray}
%\end{subequations}
Substituting the values of $Z(\varepsilon)$ and $b(\varepsilon)$ from Eqs.\eqref{eleven} and \eqref{twelve}, first two terms in Eq.\eqref{sixteen} are cancelled out and then, integration over $\varepsilon'$ yields the identity,
\begin{equation} \label{seventeen}
|a(\varepsilon)V_{1(3\epsilon)}+b(\varepsilon)V_{2(3\epsilon)}|^2 (\pi^2+|Z(\varepsilon)|^2)=1.
\end{equation}
Using Eqs.\eqref{eleven}, \eqref{twelve} and \eqref{seventeen}, we obtain the requisite probabilities of two discrete bound states,
\begin{subequations}\label{eighteen}
\begin{align}
&|a(\varepsilon)|^2=\frac{V_{1(3\epsilon)}^2\Delta_2^2}{\Delta_1^2\Delta_2^2+\pi^2(\Delta_1V_{2(3\epsilon)}^2+\Delta_2V_{1(3\epsilon)}^2)^2} \\
&|b(\varepsilon)|^2=\frac{V_{2(3\epsilon)}^2\Delta_1^2}{\Delta_1^2\Delta_2^2+\pi^2(\Delta_1V_{2\epsilon}^2+\Delta_2V_{1(3\epsilon)}^2)^2}.
\end{align}
\end{subequations}
where, $\Delta_1=\varepsilon_1-\varepsilon$ and $\Delta_2=\varepsilon_2-\varepsilon$ be the detuning of two discrete levels from the effective energy level.
\par
Finally, to obtain the Fano line-shape formula for the lambda system, we consider the dipole transition $\langle\psi(\varepsilon)|D|i\rangle$ from any arbitrary external level $|i\rangle$ to the effective (mixed) state $|\psi(\varepsilon)\rangle$. Plugging back the value of $c(\epsilon,\varepsilon)$ from Eq.\eqref{sevena}
into Eq.\eqref{fivea} and making use of Eq.\eqref{nine}, the dipole transition amplitude is given by,
%\begin{widetext}
%\begin{subequations}\label{ten}
\begin{align}\label{nineteen}
\langle\psi^\Lambda(\varepsilon)|D|i\rangle & =a(\varepsilon)\langle1|D|i\rangle+b(\varepsilon)\langle2|D|i\rangle \nonumber \\
&+a(\varepsilon)\int d{\epsilon}\langle\epsilon,3|\Big(\frac{\mathbb{P}}{\varepsilon-\epsilon}V_{1(3\epsilon)}+Z(\epsilon)V_{1(3\epsilon)}\Big)D|i\rangle \nonumber \\
&+b(\varepsilon)\int d{\epsilon}\langle\epsilon,3|\Big(\frac{\mathbb{P}}{\varepsilon-\epsilon}V_{2(3\epsilon)}+Z(\epsilon)V_{2(3\epsilon)}\Big)D|i\rangle
\end{align}
Now defining two shifted states, which are indeed an admixture of the bound and continuum states, we have,
\begin{subequations}\label{twenty}
\begin{align}
|\Phi_1\rangle &=|1\rangle+\int d\epsilon \frac{\mathbb{P}}{\epsilon-\varepsilon}V_{1(3\epsilon)}|3,\epsilon\rangle,\\
|\Phi_2\rangle &=|2\rangle+\int d\epsilon \frac{\mathbb{P}}{\epsilon-\varepsilon}V_{2(3\epsilon)}|3,\epsilon\rangle,
\end{align}
\end{subequations}
Eq.\eqref{nineteen} is reduced to
\begin{equation}\label{twentyone}
\begin{aligned}
\langle\psi^\Lambda(\varepsilon)|D|i\rangle & = a(\varepsilon)\big(\langle\Phi_1|D|i\rangle+Z(\epsilon)V_{1(3\epsilon)}\langle\varepsilon,3|D|i\rangle\big) \\
&+b(\varepsilon)\big(\langle\Phi_2|D|i\rangle+Z(\epsilon)V_{2(3\epsilon)}\langle\varepsilon,3|D|i\rangle\big).
\end{aligned}
\end{equation}
Finally to find the Fano line-shape formula
%\begin{equation}\label{tenx}
\begin{align}\label{twentytwo}
R^\Lambda_U=\Big|\frac{\langle\psi(\varepsilon)|D|i\rangle}{\langle\varepsilon,3|D|i\rangle}\Big|^2,
\end{align}
%\end{equation}
we define the line-shape indices to be,
\begin{subequations}\label{twentythree}
\begin{align}
q_1=&\frac{\langle\Phi_1 (\varepsilon)|D|i\rangle}{\pi V_{1(3\epsilon)}\langle\varepsilon,3|D|i\rangle},\\
q_2=&\frac{\langle\Phi_2(\varepsilon)|D|i\rangle}{\pi V_{2(3\epsilon)}\langle\varepsilon,3|D|i\rangle}.
\end{align}
\end{subequations}
Using Eqs.\eqref{eleven}, \eqref{twelve}, \eqref{seventeen}, $(21-23)$, the line-shape formula in Eq.\eqref{twentytwo} becomes,
\begin{equation}\label{twentyfour}
\begin{aligned}
R^\Lambda_U &=\frac{\big|\Delta_1\Delta_2+\pi q_1\Delta_1V_{2(3\epsilon)}^2+\pi q_2\Delta_2V_{1(3\epsilon)}^2\big|^2}{\Delta_1^2\Delta_2^2+(\pi\Delta_1V_{2(3\epsilon)}^2+\pi\Delta_2V_{1(3\epsilon)}^2)^2}. \\
&= \frac{\big|2\Delta_1\Delta_2+q_1\Delta_1\Gamma_{2(3\epsilon)}+ q_2\Delta_2\Gamma_{1(3\epsilon)}\big|}{4\Delta_1^2\Delta_2^2+(\Delta_1\Gamma_{2(3\epsilon)}+\Delta_2\Gamma_{1(3\epsilon)})^2}.
\end{aligned}
\end{equation}
where, $q_i (i=1,2)$ be the line-shape indices with the linewidths $\Gamma_{1(3\epsilon)}=2\pi{{V}_{1(3\epsilon)}}^2$ and $\Gamma_{2(3\epsilon)}=2\pi{{V}_{2(3\epsilon)}}^2$, respectively.
\vspace {.5cm}
\par
\subsection{\label{3b} $\Lambda$ system with middle state being continuum}
\par
For the $\Lambda$ system with middle level as the continuum state shown in Fig.~\ref{subfigure1b}, we once again plugin Eqs.\eqref{twob}, \eqref{fiveb} into Eq.\eqref{four} and obtain, (droping the symbol $\Lambda$ for notational convenience),
\begin{subequations}\label{twentyfive}
\begin{align}\label{twentyfivea}
c(\varepsilon)(\varepsilon_3-\varepsilon) + a(\varepsilon)g_{31} + \int d \epsilon b(\epsilon,\varepsilon)V_{32}(\epsilon)=0,
\end{align}
\begin{align}\label{twentyfiveb}
b(\epsilon,\varepsilon)(\epsilon-\varepsilon) + c(\varepsilon)V_{32}(\epsilon)=0,
\end{align}
\begin{align}\label{twentyfivec}
a(\varepsilon)(\varepsilon_1-\varepsilon) + c(\varepsilon)g_{31}=0,
\end{align}
\end{subequations}
respectively. Substituting the value of $b(\epsilon,\varepsilon)$ from Eq.\eqref{twentyfiveb} into Eq.\eqref{twentyfivea} and making use Dirac prescription given by Eq.\eqref{nine} we obtain the value of $Z(\epsilon)$, namely,
\begin{equation}\label{twentysix}
%\begin{align}
Z(\varepsilon)=\frac{(\varepsilon-\varepsilon_1)(\varepsilon-\varepsilon_3)-g_{13}^2}{(\varepsilon-\varepsilon_1)V_{3(2\epsilon)}^2}.
%\end{align}
\end{equation}
In deriving Eq.\eqref{twentysix}, once again we drop the principal part of the integral and assume that the interaction terms are independent of
energy, $V_{32}(\epsilon)=V_{3(2\epsilon)}$. The upper state is found to satisfiy the identity,
\begin{equation} \label{twentyseven}
|c(\varepsilon)V_{3(2\epsilon)}|^2 (\pi^2+|Z(\varepsilon)|^2)=1,
\end{equation}
while Eqs.(25-27) yield the probabilities of the bound states,
%\begin{widetext}
\begin{subequations}\label{twentyeight}
\begin{align}
&|a(\varepsilon)|^2=\frac{g_{13}^2V_{3(2\epsilon)}^2}{\pi^2\Delta_1^2V_{3(2\epsilon)}^4+(g_{13}^2-\Delta_1\Delta_3)^2}, \\
&|c(\varepsilon)|^2=\frac{V_{3(2\epsilon)}^2\Delta_{1}^2}{\pi^2\Delta_1^2V_{3(2\epsilon)}^4+(g_{13}^2-\Delta_1\Delta_3)^2},
\end{align}
\end{subequations}
%\end{widetext}
where, $\Delta_1=\varepsilon-\varepsilon_1$ and $\Delta_3=\varepsilon-\varepsilon_3$ be the detuning from the effective state defined earlier.\\
\par
To find the line-shape, similar to the previous section, we consider the dipole transition from any arbitrary level $|i>$ to the mixed state $|\psi^\Lambda>$,
\begin{subequations}\label{twentynine}
\begin{align}
\langle\psi^\Lambda(\varepsilon)|D|i\rangle & = a(\varepsilon)\langle1 |D|i\rangle + b(\varepsilon)\langle\Phi_2 |D|i\rangle +  \nonumber \\
& b(\varepsilon)Z(\epsilon)V_{3(2\epsilon)}\langle\varepsilon,3|D|i\rangle.
\end{align}
\end{subequations}
where,
\begin{equation}\label{thirty}
|\Phi_3\rangle=|3\rangle+\int d\epsilon \frac{\mathbb{P}}{\epsilon-\varepsilon}V_{3(2\epsilon)}|2,\epsilon\rangle,\\
\end{equation}
Finally defining,
\begin{subequations}\label{thirtyone}
\begin{align}\label{thirtyonea}
V_{c}=&\frac{\langle1|D|i\rangle}{\pi \langle\varepsilon,2|D|i\rangle},
\end{align}
\begin{align}\label{thirtyoneb}
q V_{3(2\epsilon)}=&\frac{\langle\Phi_3(\varepsilon)|D|i\rangle}{\pi \langle\varepsilon,2|D|i\rangle},
\end{align}
\begin{align}\label{thirtyonec}
R^\Lambda_M =& \Big|\frac{\langle\psi^\Xi(\varepsilon)|D|i\rangle}{\langle\varepsilon,2|D|i\rangle}\Big|^2,
\end{align}
\end{subequations}
and using Eqs.(27-31), the Fano line-shape formula for the lambda system for Fig.1b is given by,
\begin{equation}\label{thirtytwo}
\begin{aligned}
R^\Lambda_M & =
\frac{\pi^2}{(\pi^2+|Z(\varepsilon)|^2)}\Big|q+\frac{|Z(\varepsilon)|}{\pi}+\frac{g_{13}V_{c}}{\Delta_1V_{3(2\epsilon)}}\Big|^2 \\
& = \frac{\big|q\Delta_1\Gamma_{3(2\epsilon)}+g_{13}V_c \sqrt{2\pi\Gamma_{3(2\epsilon)}}-2q(g_{13}^2-\Delta_1\Delta_3)\big|^2}{\Delta_1^2\Gamma_{3(2\epsilon)}^2+4(g_{13}^2-\Delta_1\Delta_3)},
\end{aligned}
\end{equation}
where, $\Gamma_{3(2\epsilon)}=2\pi{{V}_{3(2\epsilon)}}^2$.
\par
\subsection{\label{3c} $\Lambda$ system with lower state being continuum}
\par
For the $\Lambda$ system with lower level continuum state shown in Fig.~\ref{subfigure1c}, we substitute Eqs.\eqref{onec}, \eqref{fivec} into Eq.\eqref{four} and obtain (droping the symbol $\Lambda$),
\begin{subequations}\label{thirtythree}
\begin{align}\label{thirtythreea}
c(\varepsilon)(\varepsilon_3-\varepsilon) + b(\varepsilon)g_{32} + \int d \epsilon a(\epsilon,\varepsilon)V_{31}(\epsilon)=0,
\end{align}
\begin{align}\label{thirtythreeb}
b(\epsilon)(\varepsilon_2-\varepsilon) + c(\varepsilon)g_{32}=0,
\end{align}
\begin{align}\label{thirtythreec}
a(\epsilon,\varepsilon)(\epsilon-\varepsilon_1) + c(\varepsilon)V_{31}(\epsilon)=0,
\end{align}
\end{subequations}
respectively. Substituting the value of $a(\epsilon,\varepsilon)$ from Eq.\eqref{thirtythreec} into Eqs.\eqref{thirtythreea} and making use Dirac prescription given by Eq.\eqref{nine} we obtain the value of $Z(\epsilon)$, namely,
\begin{equation}\label{thirtyfour}
Z(\varepsilon)=\frac{(\varepsilon-\varepsilon_2)(\varepsilon-\varepsilon_3)-g_{23}^2}{(\varepsilon-\varepsilon_2)V_{31}^2},
\end{equation}
where $V_{31}(\epsilon)=V_{3(1\epsilon)}$. Proceeding in the similar way we obtain the identity,
\begin{equation} \label{thirtyfive}
|c(\varepsilon)V_{3(1\epsilon)}|^2 (\pi^2+|Z(\varepsilon)|^2)=1,
\end{equation}
and the probabilities of the bound states are given by
\begin{subequations}\label{thirtysix}
\begin{align}
&|b(\varepsilon)|^2=\frac{g_{32}^2V_{3(1\epsilon)}^2}{\pi^2\Delta_1^2V_{3(1\epsilon)}^4+(g_{23}^2-\Delta_2\Delta_3)^2}, \\
&|c(\varepsilon)|^2=\frac{\Delta_{2}^2V_{3(1\epsilon)}^2}{\pi^2\Delta_1^2V_{3(1\epsilon)}^4+(g_{23}^2-\Delta_2\Delta_3)^2},
\end{align}
\end{subequations}
where, $\Delta_2=\varepsilon-\varepsilon_2$ and $\Delta_3=\varepsilon-\varepsilon_3$, respectively.\\
\par
To find the line-shape, similar to the previous section, we consider the dipole transition from any arbitrary level $|i>$ to the mixed state $|\psi^\Lambda>$,
\begin{equation}\label{thirtyseven}
\begin{aligned}
\langle\psi^\Lambda(\varepsilon)|D|i\rangle=&a(\varepsilon)\langle1|D|i\rangle + c(\varepsilon)\langle\Phi_2 |D|i\rangle\\
+&b(\varepsilon)Z(\epsilon)V_{3(1\epsilon)}\langle\varepsilon,3|D|i\rangle.
\end{aligned}
\end{equation}
where,
\begin{equation}\label{thirtyeight}
|\Phi_3(\varepsilon)\rangle=|3\rangle+\int d\epsilon \frac{\mathbb{P}}{\epsilon-\varepsilon}V_{3(1\epsilon)}|1,\epsilon\rangle.
\end{equation}
Finally defining,
\begin{subequations}\label{thirtynine}
\begin{align}
V_{c}=&\frac{\langle1|D|i\rangle}{\pi\langle\varepsilon,1|D|i\rangle},\\
q V_{3(1\epsilon)}=&\frac{\langle\Phi_3(\varepsilon)|D|i\rangle}{\pi\langle\varepsilon,1|D|i\rangle},\\
R^\Lambda_L =& \Big|\frac{\langle\psi^\Xi(\varepsilon)|D|i\rangle}{\langle\varepsilon,1|D|i\rangle}\Big|^2,
\end{align}
\end{subequations}
and using Eqs.(27-30), the Fano line-shape formula for the lambda system with lower state continum is given by,
\begin{equation}\label{forty}
\begin{aligned}
R^\Lambda_L & =
 \frac{\pi^2}{(\pi^2+|Z(\varepsilon)|^2)}\Big|q+\frac{|Z(\varepsilon)|}{\pi}+\frac{g_{23}V_{c}}{\Delta_1V_{3(1\epsilon)}}\Big|^2 \\
& = \frac{q\Delta_2\Gamma_{3(1\epsilon)}+\big|g_{23}V_c \sqrt{2\pi\Gamma_{3(1\epsilon)}}-2q(g_{23}^2-\Delta_2\Delta_3)\big|^2}{\Delta_2^2\Gamma_{3(1\epsilon)}^2+4(g_{23}^2-\Delta_2\Delta_3)},
\end{aligned}
\end{equation}
where, $\Gamma_{3(1\epsilon)}=2\pi{{V}_{3(1\epsilon)}}^2$. The evaluation of the line-shapes of the cascade and vee systems are similar and in the next section here we only quote the results.
\par
\section{\label{4} $\Xi$ system}
\par
\subsection{\label{4a} $\Xi$ system with upper state being continuum}
\par
For the $\Xi$ type system shown in Fig.~\ref{subfigure2a}, the probabilities of the upper and lower levels are given by,
\begin{subequations}\label{fortyone}
\begin{align}
&|a(\varepsilon)|^2=\frac{g_{12}^2V_{2(3\epsilon)}^2}{\pi^2\Delta_1^2V_{2(3\epsilon)}^4+(\Delta_1\Delta_2-g_{12}^2)^2}, \\
&|b(\varepsilon)|^2=\frac{\Delta_1^2V_{2(3\epsilon)}^2}{\pi^2\Delta_1^2V_{2(3\epsilon)}^4+(\Delta_1\Delta_2-g_{12}^2)^2},
\end{align}
\end{subequations}
where, $\Delta_1=\varepsilon-\varepsilon_1$ and $\Delta_2=\varepsilon-\varepsilon_2$, respectively.
The probability amplitude of the middle state satisfies the identity,
\begin{equation} \label{fortytwo}
|b(\varepsilon)V_{2(3\epsilon)}|^2 (\pi^2+|Z(\varepsilon)|^2)=1,
\end{equation}
where the value of $Z(\varepsilon)$ is given by,
\begin{equation}\label{fortythree}
Z(\varepsilon)=\frac{\Delta_1\Delta_2-g_{12}^2}{\Delta_1V_{2(3\epsilon)}^2}.
\end{equation}
Thus the Fano line-shape of the cascade system with upper level continuum is given by,
\begin{equation}\label{fortyfour}
\begin{aligned}
R^\Xi_U & =\frac{\pi^2}{(\pi^2+|Z(\varepsilon)|^2)}\Big|q+\frac{|Z(\varepsilon)|}{\pi}+\frac{g_{12}V_c}{\Delta_1V_{2(3\epsilon)}}\Big|^2,\\
&=\frac{\big|q \Delta_{1}\Gamma_{2(3\epsilon)}+g_{12}V_c \sqrt{2\pi\Gamma_{2(3\epsilon)}}-2(g_{13}^2-\Delta_1\Delta_3)\big|^2}{\Delta_1^2\Gamma_{2(3\epsilon)}^2+4(g_{12}^2-\Delta_1\Delta_2)^2},
\end{aligned}
\end{equation}
where, $\Gamma_{2(3\epsilon)}=2\pi{{V}_{2(3\epsilon)}}^2$.

\subsection{\label{4b} $\Xi$ system with middle state being continuum}
\par
For the $\Xi$ system with the middle state as the continuum state shown in Fig.~\ref{subfigure2b}, the probabilities of the lower and upper levels are given by,
\begin{subequations}\label{fortyfive}
\begin{align}
&|a(\varepsilon)|^2=\frac{\Delta_3^2V_{1(2\epsilon)}^2}{\pi^2(\Delta_1^2V_{3(2\epsilon)}^2+\Delta_3V_{1(2\epsilon)}^2)^2+\Delta_1^2\Delta_3^2}, \\
&|c(\varepsilon)|^2=\frac{\Delta_1^2V_{3(2\epsilon)}^2}{\pi^2(\Delta_1^2V_{3(2\epsilon)}^2+\Delta_3V_{1(2\epsilon)}^2)^2+\Delta_1^2\Delta_3^2},
\end{align}
\end{subequations}
where, $\Delta_1=\varepsilon-\varepsilon_1$ and $\Delta_3=\varepsilon-\varepsilon_3$, respectively. Similar to previous Section, the probability amplitudes of these two bound states  satisfy,
\begin{equation} \label{fortysix}
(|a(\varepsilon)V_{1(2\epsilon)}+c(\varepsilon)V_{3(2\epsilon)}|^2)(\pi^2+|Z(\varepsilon)|^2)=1,
\end{equation}
where the value of $Z(\varepsilon)$ is given by,
\begin{equation}\label{fortyseven}
Z(\varepsilon)=\frac{\Delta_1\Delta_3}{\Delta_1V_{3(2\epsilon)}^2+\Delta_3V_{1(2\epsilon)}^2}.
\end{equation}
Thus the Fano line-shape for such system is found to be,
\par
\begin{equation}\label{fortyseight}
\begin{aligned}
R^\Xi_M & =\frac{\big|\Delta_1\Delta_3+\pi q_1\Delta_3V_{1(2\epsilon)}^2+\pi q_3\Delta_1V_{3(2\epsilon)}^2\big|^2}{\Delta_1^2\Delta_3^2+\pi^2(\Delta_3V_{1(2\epsilon)}^2+\Delta_1V_{3(2\epsilon)}^2)^2}\\
&= \frac{\big|2\Delta_1\Delta_3+q_1\Delta_3\Gamma_{1(2\epsilon)}+ q_3\Delta_1\Gamma_{3(2\epsilon)}\big|^2}{4\Delta_1^2\Delta_3^2+(\Delta_3\Gamma_{1(2\epsilon)}+\Delta_1\Gamma_{3(2\epsilon)})^2},
\end{aligned}
\end{equation}
where, $q_i (i=1,3)$ be the line-shape indices with the linewidths $\Gamma_{1(2\epsilon)}=2\pi{{V}_{1(2\epsilon)}}^2$ and $\Gamma_{3(2\epsilon)}=2\pi{{V}_{3(2\epsilon)}}^2$, respectively.
\subsection{\label{4c} $\Xi$ system with lower state being continuum}
\par
For the $\Xi$ system with the lower state continuum shown in Fig.~\ref{subfigure2c}, the probabilities of middle and upper states are given by,
\begin{subequations}\label{fortynine}
\begin{align}
&|b(\varepsilon)|^2=\frac{\Delta_1^2V_{2(1\epsilon)}^2}{\pi^2\Delta_3^2V_{2(1\epsilon)}^4+(\Delta_2\Delta_3-g_{23}^2)^2},\\
&|c(\varepsilon)|^2=\frac{g_{23}^2V_{2(1\epsilon)}^2}{\pi^2\Delta_3^2V_{2(1\epsilon)}^4+(\Delta_2\Delta_3-g_{23}^2)^2},
\end{align}
\end{subequations}
where, $\Delta_2=\varepsilon-\varepsilon_2$ and $\Delta_3=\varepsilon-\varepsilon_3$, respectively.
The probability amplitude of the middle state follows,
\begin{equation} \label{fifty}
|b(\varepsilon)V_{2(1\epsilon)}|^2 (\pi^2+|Z(\varepsilon)|^2)=1,
\end{equation}
where the value of $Z(\varepsilon)$ is given by,
\begin{equation}\label{fiftyone}
%\begin{align}
Z(\varepsilon)=\frac{\Delta_2\Delta_3-g_{23}^2}{\Delta_3V_{2(1\epsilon)}^2}.
%\end{align}
\end{equation}
Thus the Fano line-shape of the cascade system with the middle level as the continuum state is given by,
\begin{equation}\label{fiftytwo}
\begin{aligned}
R^\Xi_L & =\frac{\pi^2}{(\pi^2+|Z(\varepsilon)|^2)}\Big|q+\frac{|Z(\varepsilon)|}{\pi}+r_{L}^\Xi\frac{g_{23}V_c}{\Delta_3V_{2(1\epsilon)}}\Big|^2, \\
&=\frac{\big|q \Delta_{3}\Gamma_{2(1\epsilon)}+g_{23}V_c \sqrt{2\pi\Gamma_{2(1\epsilon)}}-2(g_{23}^2-\Delta_2\Delta_3)\big|^2}{\Delta_3^2\Gamma_{2(1\epsilon)}^2+4(g_{23}^2-\Delta_2\Delta_3)^2},
\end{aligned}
\end{equation}
where, $\Gamma_{2(1\epsilon)}=2\pi{{V}_{2(1\epsilon)}}^2$.
\par
\section{\label{5} $V$ system}
\par
\subsection{\label{5a} $V$ system with upper state being continuum}
\par
The probabilities of the middle and lower states for the $V$ system with upper state as the continuum state (Fig.~\ref{subfigure3a}) are,
\begin{subequations}\label{fiftythree}
\begin{align}
&|b(\varepsilon)|^2=\frac{g_{12}^2V_{1(3\epsilon)}^2}{\pi^2\Delta_2^2V_{1(3\epsilon)}^4+(\Delta_1\Delta_2-g_{12}^2)^2},\\
&|a(\varepsilon)|^2=\frac{\Delta_2^2V_{1(3\epsilon)}^2}{\pi^2\Delta_2^2V_{1(3\epsilon)}^4+(\Delta_1\Delta_2-g_{12}^2)^2},
\end{align}
\end{subequations}
where, $\Delta_1=\varepsilon-\varepsilon_1$ and $\Delta_2=\varepsilon-\varepsilon_2$, respectively. The probability amplitude of the lower state  satisfies,
\begin{equation} \label{fiftyfour}
|c(\varepsilon)V_{1(3\epsilon)}|^2 (\pi^2+|Z(\varepsilon)|^2)=1,
\end{equation}
where the value of $Z(\varepsilon)$ is given by,
\begin{equation}\label{fiftyfive}
Z(\varepsilon)=\frac{\Delta_1\Delta_2-g_{12}^2}{\Delta_2V_{1(3\epsilon)}^2}.
\end{equation}
Thus the Fano line-shape of the vee system with upper level continuum is found to be,
\begin{equation}\label{fiftysix}
\begin{aligned}
R^V_U & =\frac{\pi^2}{(\pi^2+|Z(\varepsilon)|^2)}\Big|q+\frac{|Z(\varepsilon)|}{\pi}+\frac{g_{12}V_c}{\Delta_2V_{1(3\epsilon)}}\Big|^2,\\
&=\frac{\big|q \Delta_{2}\Gamma_{1(3\epsilon)}+g_{12}V_c \sqrt{2\pi\Gamma_{1(3\epsilon)}}-2(g_{12}^2-\Delta_1\Delta_2)\big|^2}{\Delta_2^2\Gamma_{1(3\epsilon)}^2+4(g_{12}^2-\Delta_1\Delta_3)^2}.
\end{aligned}
\end{equation}
where, $\Gamma_{1(3\epsilon)}=2\pi{{V}_{1(3\epsilon)}}^2$.

\subsection{\label{4b} $V$ system with middle state being continuum}
\par
For the $V$ system with the middle state as the continuum state (Fig.~\ref{subfigure3b}), the probabilities of the lower and upper states are given by,
\begin{subequations}\label{fiftyseven}
\begin{align}
&|a(\varepsilon)|^2=\frac{\Delta_3^2V_{1(2\epsilon)}^2}{\pi^2\Delta_3^2V_{1(2\epsilon)}^4+(\Delta_1\Delta_3-g_{13}^2)^2},\\
&|c(\varepsilon)|^2=\frac{g_{13}^2V_{1(3\epsilon)}^2}{\pi^2\Delta_3^2V_{1(2\epsilon)}^4+(\Delta_1\Delta_3-g_{13}^2)^2},
\end{align}
\end{subequations}
where, $\Delta_1=\varepsilon-\varepsilon_1$ and $\Delta_3=\varepsilon-\varepsilon_3$, respectively.
The probability amplitude of the lower state satisfies,
\begin{equation} \label{fiftyeight}
|a(\varepsilon)V_{1(2\epsilon)}|^2(\pi^2+|Z(\varepsilon)|^2)=1,
\end{equation}
where the value of $Z(\varepsilon)$ is given by,
\begin{equation}\label{fiftynine}
Z(\varepsilon)=\frac{\Delta_1\Delta_3-g_{13}^2}{\Delta_3V_{1(2\epsilon)}^2}.
\end{equation}
Thus the Fano line-shape of the requisite vee system is found to be,
\begin{equation}\label{sixty}
\begin{aligned}
R^V_M & =\frac{\pi^2}{(\pi^2+|Z(\varepsilon)|^2)}\Big|q+\frac{|Z(\varepsilon)|}{\pi}+\frac{g_{12}V_c}{\Delta_3V_{1(2\epsilon)}}\Big|^2,\\
&=\frac{\big|g_{12}V_c \sqrt{2\pi\Gamma_{1(3\epsilon)}}+q \Delta_{2}\Gamma_{1(3\epsilon)}-2(g_{12}^2-\Delta_1\Delta_2)\big|^2}{\Delta_2^2\Gamma_{1(3\epsilon)}^2+4(g_{12}^2-\Delta_1\Delta_3)^2},
\end{aligned}
\end{equation}
where, $\Gamma_{1(2\epsilon)}=2\pi{{V}_{1(2\epsilon)}}^2$.
\par
\subsection{\label{4c} $V$ system with lower state being continuum}
\par
The probabilities of the upper and middle states for the $V$ system with the lower state as the continuum (Fig.~\ref{subfigure3c}) are given by,
\begin{subequations}\label{sixtyone}
\begin{align}
&|c(\varepsilon)|^2=\frac{\Delta_2^2V_{3(1\epsilon)}^2}{\pi^2(\Delta_2^2V_{3(1\epsilon)}^2+\Delta_3V_{2(1\epsilon)}^2)^2+\Delta_1^2\Delta_3^2},\\
&|b(\varepsilon)|^2=\frac{\Delta_3^2V_{2(1\epsilon)}^2}{\pi^2(\Delta_2^2V_{3(1\epsilon)}^2+\Delta_3V_{2(1\epsilon)}^2)^2+\Delta_1^2\Delta_3^2},
\end{align}
\end{subequations}
where, $\Delta_2=\varepsilon-\varepsilon_2$ and $\Delta_3=\varepsilon-\varepsilon_3$, respectively.
The probability amplitudes of the bound states, namely, the middle and upper states satisfy the identity,
\begin{equation} \label{sixtytwo}
|b(\varepsilon)V_{2(1\epsilon)}+c(\varepsilon)V_{3(1\epsilon)}|^2(\pi^2+|Z(\varepsilon)|^2)=1,
\end{equation}
where the value of $Z(\varepsilon)$ is given by,
\begin{equation}\label{sixtythree}
Z(\varepsilon)=\frac{\Delta_2\Delta_3}{\Delta_3V_{2(1\epsilon)}^2+\Delta_2V_{3(1\epsilon)}^2}.
\end{equation}
and the Fano line-shape of the vee system for Fig.3c is found to be,
\begin{equation}\label{sixtyfour}
\begin{aligned}
R^V_L &=\frac{\big|\Delta_2\Delta_3+\pi q_2\Delta_3V_{2(1\epsilon)}^2+\pi q_3\Delta_2V_{3(1\epsilon)}^2\big|^2}{\Delta_2^2\Delta_3^2+\pi^2(\Delta_3V_{2(1\epsilon)}^2+\Delta_2V_{3(1\epsilon)}^2)^2}.\\
&= \frac{\big|2\Delta_2\Delta_3+q_2\Delta_3\Gamma_{2(1\epsilon)}+ q_2\Delta_2\Gamma_{3(1\epsilon)}\big|^2}{4\Delta_2^2\Delta_3^2+(\Delta_3\Gamma_{2(1\epsilon)}+\Delta_2\Gamma_{3(1\epsilon)})^2},
\end{aligned}
\end{equation}
where, $q_i (i=2,3)$be the line-shape indices with the linewidths $\Gamma_{2(1\epsilon)}=2\pi{{V}_{2(1\epsilon)}}^2$ and $\Gamma_{3(1\epsilon)}=2\pi{{V}_{3(1\epsilon)}}^2$, respectively.
%%%%%%%%%%%%%%%%%%%%%%%%%%%%%%%%%%%%%%%%%%%%%%%%%

\section{\label{3x} Numerical Results}
\par
Before discussing the line-shapes of all configurations, let us note the distinctive features of the proposed three-level-like continuum models characterized by two detuning parameters $\Delta_{i}$ and $\Delta_{j}$ ($i, j=1,2, 3$ and $i\neq j$) along with one or more lineshape indices $q_{i}$. In particular, we note that the systems with two bound-continuum transitions as shown in Figs.1a, 2b and 3c are characterized by two different Fano asymmetry parameters $q_{i}$ and $q_{j}$,  while the remaining systems shown in Figs.1b,c, Figs.2a,c and Figs.3a,b with one bound-continuum transition have single asymmetry parameter $q$, respectively. Let us first consider the line-shape formula of the lambda system given by Eq.\eqref{twentyfour}, \eqref{thirtytwo} and \eqref{forty}, respectively. Fig.4a shows the variation of the line-shape in Eq.\eqref{twentyfour} with the detuning frequency $\Delta_1$ for different constant values of $(\Delta_2, q_2)$. We note that, even for $q_1=0$ and $q_{2}=0$ (colour black), the Fano profiles are no longer symmetric due to the interference between two channels and the asymmetry grows with the increase of $q_{2}$. Such Fano pattern of the plot of Eqs.\eqref{thirtytwo} and \eqref{forty} displayed in Fig.(4b,c), which are characterized by one asymmetry parameter $q$, remains unchanged.

\begin{figure}[ht]
%\begin{minipage}{\textwidth}
\includegraphics[width=16.00cm,height=10.25cm,keepaspectratio]{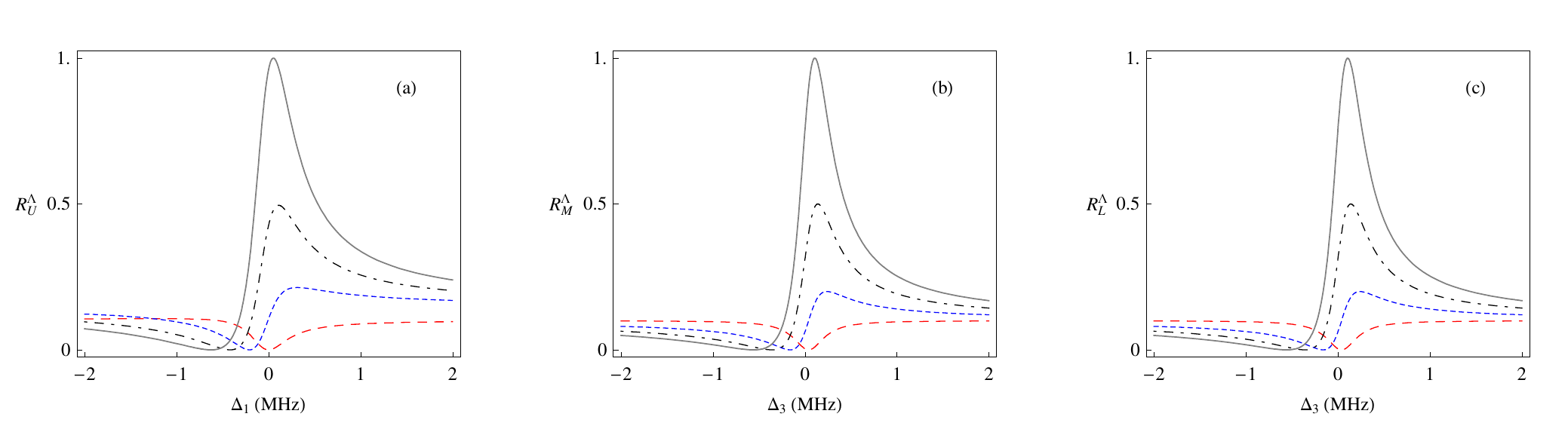}
%\includegraphics[]{lambda_fano.eps}
%\begin{flushleft}
\caption{\label{lambda-fano}: Typical Fano profile of $\Lambda$-type system from a) Eq.(24) with $\Gamma_{1(3\epsilon)}=.5$ MHz, $\Gamma_{2(3\epsilon)}=.4$ MHz, $\Delta_{2}=1$MHz, for $q_{1}=0$, $q_{2}=0$ (Red, Dashed) and for $q_{1}=1$, $q_{2}=1$ (Blue,dotted), $2$ (black, dot-dashed), $3$ (Grey, Solid); b) Eq.(32) with $g_{13}= .2$ MHz, $\Gamma_{3(2\epsilon)}=.4$ MHz, $V_c=.1$, $\Delta_{1}=1$MHz, for $q=0$ (Red, Dashed), $1$ (Blue,dotted), $2$ (black, dot-dashed), $3$ (Grey, Solid); c) Eq.(40) for $g_{23}=.2$ MHz, $\Gamma_{3(1\epsilon)}=.4$MHz, $V_c=0.1$, $\Delta_{2}=1$MHz, for $q=0$ (Red, Dashed), $1$ (Blue, dotted), $2$ (black, dot-dashed), $3$ (Grey, Solid), respectively.}
%\end{flushleft}
%\end{minipage}
\end{figure}
%\end{widetext}
%\begin{widetext}
\begin{figure}[ht]
\begin{minipage}{\textwidth}
\includegraphics[width=16.00cm,height=10.25cm,keepaspectratio]{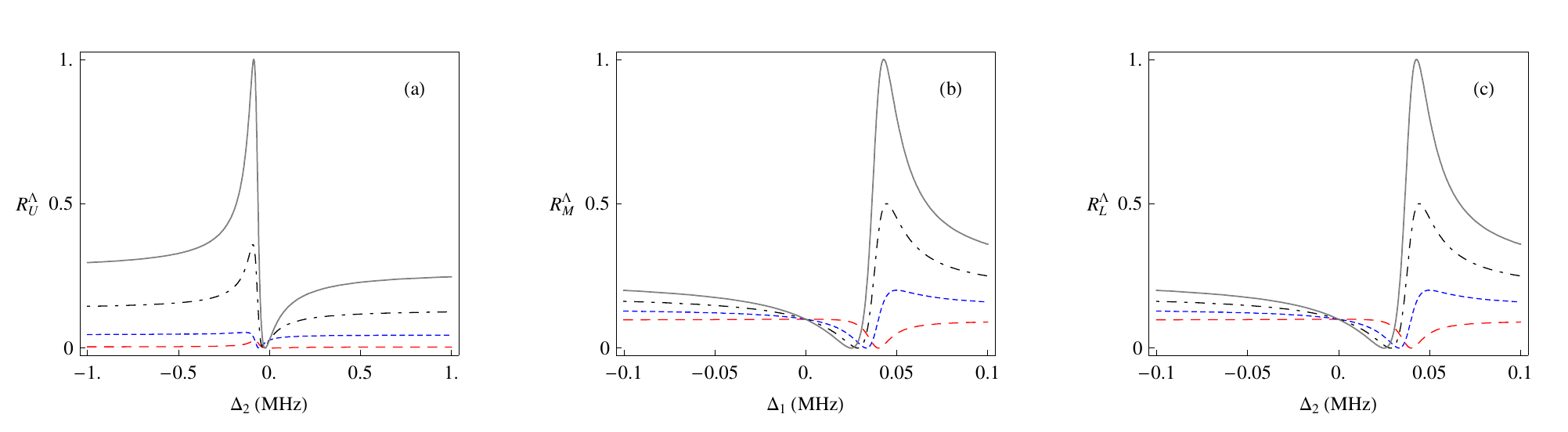}
\begin{flushleft}
\caption{\label{modified-lambda-fano}: Modified Fano profile of $\Lambda$-type system from a) Eq.(24) $\Gamma_{1(3\epsilon)}=.5$ MHz, $\Gamma_{2(3\epsilon)}=.4$ MHz, $\Delta_{1}=.1$MHz, $q_{2}=0$, $q_{1}=0$ (Red, Dashed) and for $q_{1}=1$, $q_{2}=1$ (Blue, dotted), $2$ (Red, Dashed, dot-dashed), $3$ (Grey, Solid); b) Eq.(32) with $g_{13}= 0.2$ MHz, $\Gamma_{3(2\epsilon)}=.4$ MHz, $V_c=.2$, $\Delta_{3}=1$ MHz, $q=0$ (Black), $1$ (Blue, dotted), $2$ (black, dot-dashed), $3$ (Grey, Solid); c) Eq.(40) with $g_{23}=.2$ MHz, $\Gamma_{3(1\epsilon)}= .4$MHz, $V_c=.1$ MHz, $\Delta_{3}=1$ MHz, $q=0$ (Red, Dashed), $1$(Blue, dotted), $2$ (black, dot-dashed), $3$ (Grey, Solid), respectively.}
\end{flushleft}
\end{minipage}
\end{figure}
%\end{widetext}
%\begin{widetext}
\begin{figure}[ht]
\begin{minipage}{\textwidth}
\includegraphics[width=16.00cm,height=10.25cm,keepaspectratio]{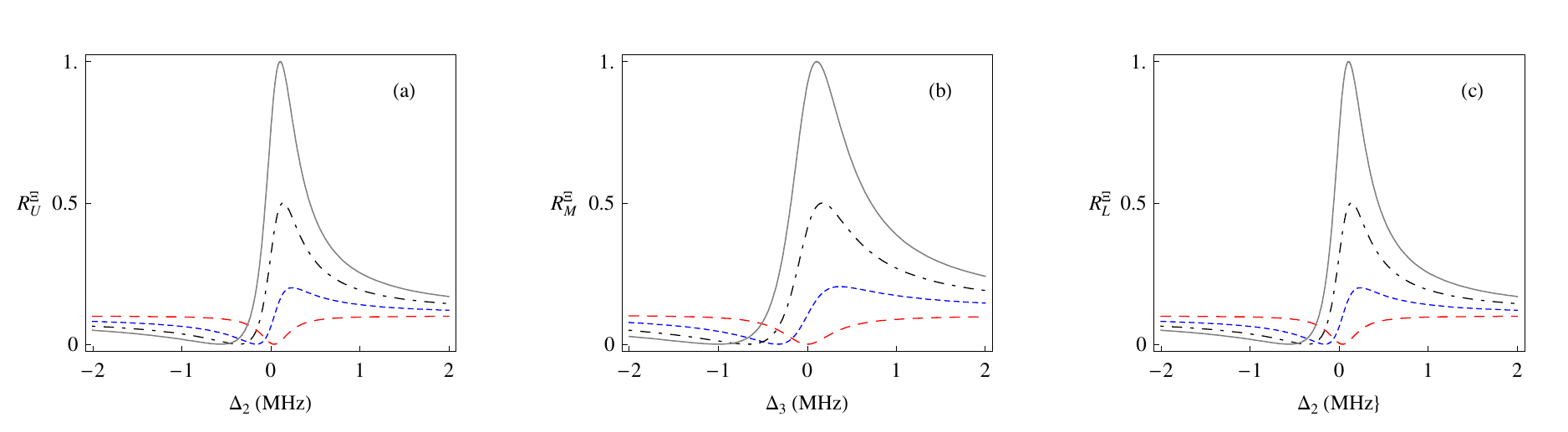}
\begin{flushleft}
\caption{\label{cascade-fano}: Typical Fano profile of $\Xi$-type system from a) Eq.(44) with $g_{12}=.2$ MHz, $\Gamma_{2(3\epsilon)}=.4$ MHz, $V_c =.1$ $\Delta_{1}=1$ MHz, for $q =0$ (Red, Dashed), $1$ (Blue, dotted), $2$ (black, dot-dashed), $3$ (Grey, Solid); b) Eq.(48) with $\Gamma_{1(2\epsilon)}= .1$ MHz, $\Gamma_{3(2\epsilon)}= 0.7$ MHz, $\Delta_{1}=1$ MHz, for $q_1=0$, $q_3=0$ (Red, Dashed) and for $q_{1}=1$, $q_{3}=1$, (Blue, dotted) $2$ (Green), $3$ (Grey, Solid); c) Eq.(52) with $g_{23}=.2$ MHz, $\Gamma_{2(1\epsilon)}=.4$ MHz, $V_c=.1$ MHz, $\Delta_{3}=1$ MHz, for $q=0$ (Red, Dashed), $1$(Blue, dotted), $2$ (black, dot-dashed), $3$ (Grey, Solid) respectively.}
\end{flushleft}
\end{minipage}
\end{figure}
\begin{figure}[ht]
\begin{minipage}{\textwidth}
\includegraphics[width=16.00cm,height=10.25cm,keepaspectratio]{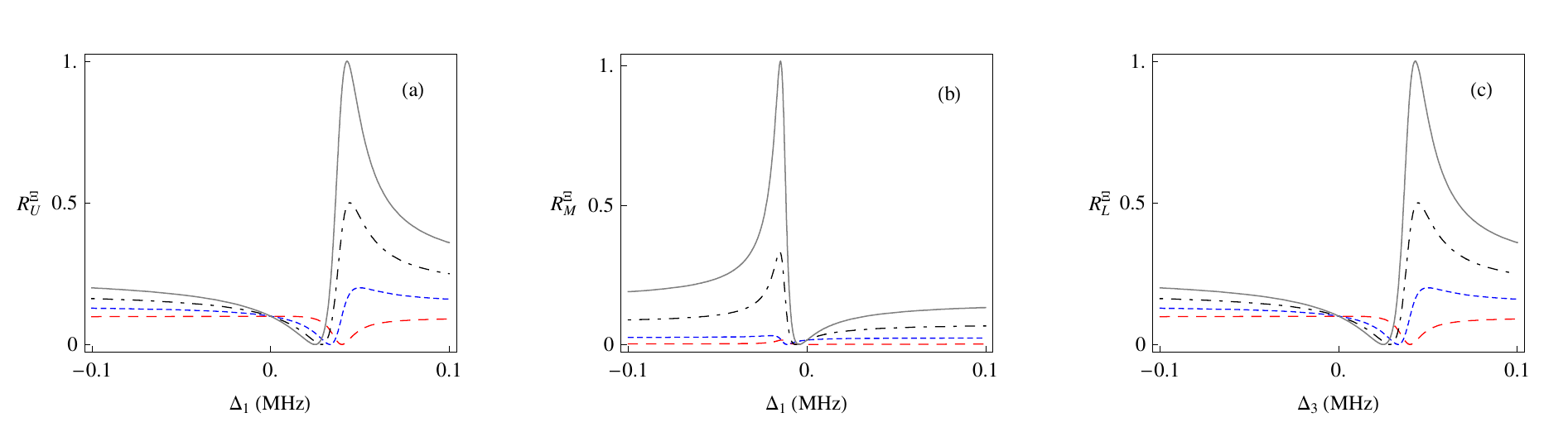}
\begin{flushleft}
\caption{\label{modified-cascade-fano}: Modified Fano profile of $\Xi$-type system from a) Eq.(40) with $g_{12}=.2$ MHz, $\Gamma_{2(3\epsilon)}=  .4$ MHz, $V_c = .1$ MHz $\Delta_{2}=1$ MHz, for $q =0$ (Red, Dashed), $1$ (Blue, dotted), $2$ (black, dot-dashed), $3$ (Grey, Solid); b) Eq.(48) with $\Gamma_{1(2\epsilon)}=.1$ MHz, $\Gamma_{3(2\epsilon)}=.7$ MHz, $\Delta_{3}=.1$ MHz, for $q_3=0$, $q_1=0$ (Red, Dashed) and for $q_{1}=1$, $q_{3}=1$, (Blue, dotted), $2$ (black, dot-dashed), $3$ (Grey, Solid); c) Eq.(52) for $g_{23}= .2$ MHz, $\Gamma_{2(1\epsilon)}=.4$MHz, $V_c=.1$, $\Delta_{2}=1$ MHz, for $q=0$ (Red, Dashed), $1$(Blue, dotted), $2$ (black, dot-dashed), $3$ (Grey, Solid), respectively.}
\end{flushleft}
\end{minipage}
\end{figure}
\begin{figure}[ht]
\begin{minipage}{\textwidth}
\includegraphics[width=16.00cm,height=10.25cm,keepaspectratio]{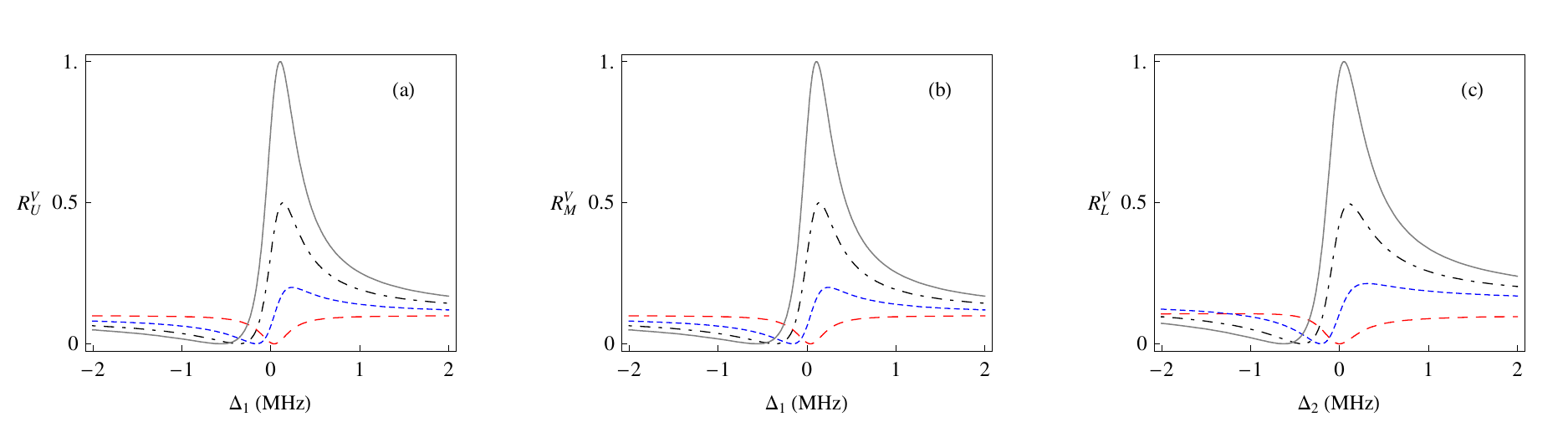}
\begin{flushleft}
\caption{\label{vee-fano}: Typical Fano profile of $V$-type system from a) Eq.(56) $g_{12}=.2$MHz, $\Gamma_{1(3\epsilon)}=.4$ MHz, $V_c =.1$  and $\Delta_{2}=1$MHz, for $q =0$ (Red, Dashed), $1$ (Blue, dotted), $2$ (black, dot-dashed), $3$ (Grey, Solid); b) Eq.(60) with $g_{13}= 0.2$ MHz, $\Gamma_{1(2\epsilon)}=.4$ MHz, $V_c=.1$ $\Delta_{3}=1$MHz, for $q = 0$ (Red, Dashed), $1$ (Blue, dotted), $2$, (black, dot-dashed), $3$ (Grey, Solid); c) Eq.(64) for $\Gamma_{2(1\epsilon)}=.5$ MHz, $\Gamma_{3(1\epsilon)}=.4$MHz, $\Delta_{3}=1$ MHz, for $q_2=0$, $q_3=0$ (Red, Dashed) and for $q_{3}=1$, $q_{2}=1$, (Blue, dotted), $2$ (black, dot-dashed), $3$ (Grey, Solid), respectively.}
\end{flushleft}
\end{minipage}
\end{figure}
\begin{figure}[ht]
\begin{minipage}{\textwidth}
\includegraphics[width=16.00cm,height=10.25cm,keepaspectratio]{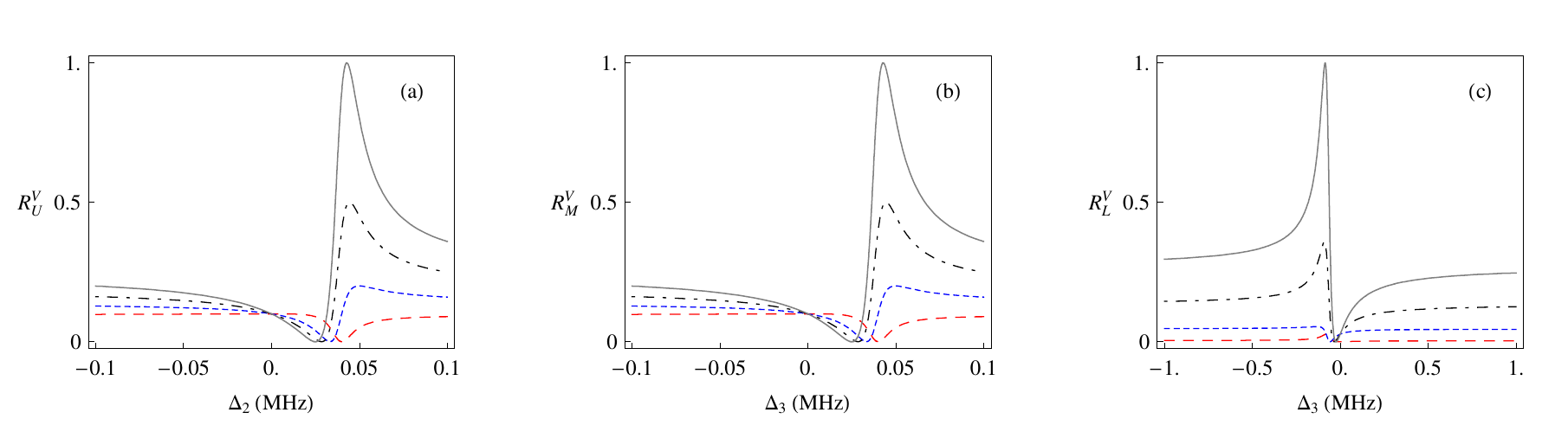}
\begin{flushleft}
\caption{\label{vee-modified-fano}: Modified Fano profile of $V$-type system from i) Eq.(56) with $g_{12}=.2$ MHz, $\Gamma_{1(3\epsilon)}=.4$MHz, $V_c =.2$MHz $\Delta_{1}=1$ MHz, for $q =0$ (Red, Dashed), $1$ (Blue, dotted), $2$ (black, dot-dashed), $3$ (Grey, Solid); ii) Eq.(60) with $g_{13}=.2$MHz, $\Gamma_{1(2\epsilon)}=.4$ MHz, $V_c=.1$ MHz, $\Delta_{1}=1$ MHz, for $q=0$  (Red, Dashed), $1$ (Blue, dotted),$2$, (black, dot-dashed), $3$ (Grey, Solid); iii) Eq.(64) with $\Gamma_{2(1\epsilon)}=.5$MHz, $\Gamma_{3(1\epsilon)}=.4$MHz, $\Delta_{2}=.1$MHz, for $q_2=0$, $q_3=0$ (Red, Dashed) and for $q_{3}=1$, $q_{2}=1$ (Blue, dotted), $2$ (black, dot-dashed), $3$ (Grey, Solid), respectively.}
\end{flushleft}
\end{minipage}
\end{figure}
\par
On the other hand, if we consider the plot of Eq.\eqref{twentyfour} for different values of $\Delta_2$ with fixed $(\Delta_1, q_2)$, i.e., by interchanging the detuning frequencies only, we obtain a modified line-shape of the lambda system shown in Fig.5(a). For such system with two bound-to-continuum transition, at resonance  $\Delta_1=0$, the lineshape is independent of the asymmetric factor except for $q_1=0=q_2$. In sharp contrast, for the remaining plots with one bound-continuum system shown in Fig.5(b) and Fig.5(c), at resonance $\Delta_i=0$ ($i=1,2$), all lineshapes are independent the asymmetric factor $q$.
Proceeding in similar way, we obtain Fano and modified Fano plots of the $\Xi$ and $V$ systems shown in Fig.12-13 and Fig.14-15, respectively and the behaviour of all configurations is similar to that of the $Lambda$ system. In our numerical plots, we have taken abitrary values of the parameters, keeping in mind the realistic range of parameters applicable to optical dipole transitions in atomic and molecular physics. Since our models are applicable to systems with small or negligible relaxations or damping, one needs to find atomic or molecular levels with narrow-line transitions, with linewidth typically below one MHz. With the recent advent of high precision spectroscopy of photoassociation at the inetrface of atomic and molecular systems, particularly with narrow-line intercombination transitions, the experimental realisations of our the models discussed here seem to be quite promising.

\section{\label{3x} Conclusions and outlook}
The purpose of this paper is to develop a comprehensive Fano-inspired model of three-level-like continuum system. It is shown that the minimal extension of a generic three-level system by replacing its one bound state with a continuum of states leads to nine possible distinct configurations. The line-shape of each model is studied for two different detuning frequencies which, in addition to typical asymmetric Fano plot, gives rise to a modified spectral features depending upon the system parameters. The existence of two different detuning frequencies, which may be identified with the pump or probe modes, is a natural outcome of our analysis and is extremely important to develop a suitable control mechanism of dispersion and absorption in three-level-like  continuum systems. The generalization of the proposed continuum models incorporating dissipation in Lindblad or in other ways, may form the basis of studying wide range of quantum optical phenomena including continuum based model of EIT.

\section*{Acknowledgement}
S.S. thanks Department of Science and Technology, New Delhi for partial support. S.S. and T.K.D. also thank Raman Center for Atomic, Molecular and Optical Sciences (CAMOS), Indian Association for the Cultivation of Science, Kolkata, for hospitality.

\bibliographystyle{tfp}
\bibliography{fano3}

\end{document}